\documentclass[useAMS,usenatbib,usegraphicx]{mn2e}
\usepackage{graphicx}
\usepackage{natbib}
\bibpunct{(}{)}{;}{a}{}{,}
\usepackage{journals}   

\begin{document}

\title[Using radio stars for GCRF--ICRF link]{Using radio stars to link the {\it Gaia} and VLBI reference frames}
\author[Z. Malkin]{Zinovy Malkin$^{1,2,3}$\\
  $^1$Pulkovo Observatory, St.~Petersburg, 196140, Russia\\
  $^2$St.~Petersburg State University, St.~Petersburg, 199034, Russia\\
  $^3$Kazan Federal University, Kazan, 420000, Russia}
\maketitle

\maketitle

\begin{abstract}
A possible method for linking the optical {\it Gaia} Celestial Reference Frame (GCRF) to the VLBI-based International Celestial Reference Frame
(ICRF) is to use radio stars in a manner similar to that in the linking of the {\it Hipparcos} Celestial Reference Frame (HCRF) to ICRF.
In this work, an obtainable accuracy of the orientation angles between GCRF and ICRF frames was estimated by Monte Carlo simulation.
If the uncertainties in the radio star positions obtained by VLBI are in the range of 0.1--4~mas and those obtained by {\it Gaia} are
in the range of 0.005--0.4~mas, the orientation angle uncertainties are 0.018--0.72~mas if 46 radio stars are used,
0.013--0.51~mas if 92 radio stars are used, and 0.010--0.41~mas if 138 radio stars are used.
The general conclusion from this study is that a properly organized VLBI programme for radio star observation with a reasonable load
on the VLBI network can allow for the realization of GCRF--ICRF link with an error of about 0.1~mas.
\end{abstract}

\begin{keywords}
astrometry -- reference systems -- techniques: interferometric
\end{keywords}

\maketitle


\section{Introduction}

Establishing the link (orientation angles) between radio and optical reference frames is a key problem in fundamental astrometry in the coming years.
Since 1998, IAU-adopted International Celestial Reference System (ICRS) is represented by VLBI-based International Celestial Reference Frame
(ICRF, \citet{Ma1998}.
ICRF2 is the current IAU-approved official ICRF realization \citep{Fey2015}.
The next ICRF realization, ICRF3, is expected to be completed by 2018 \citep{Jacobs2014}.
The new-generation optical {\it Gaia} Celestial Reference Frame (GCRF) from the {\it Gaia} mission is targeted to be available in 2021,
with the intermediate release expected around 2018 \citep{Lindegren2008,Mignard2015}.
Each of these two frames is anticipated to provide a position accuracy of several tens of $\mu$as.
These two highly accurate frames should be linked to each other to provide a consistent astrometric multi-wave ICRS realization.
This ICRF--GCRF link is supposed to be achieved using common ICRF--GCRF extragalactic objects observed at both radio and optical wavelengths
\citep{Bourda2008,Taris2013,Zacharias2014,LeBail2016}.

However, the accuracy of the link between ICRF and GCRF based on extragalactic objects is limited by several factors.
These factors include the different locations of radio and optical brightness centroids; radio source structure, which is generally variable;
core-shift effect; and asymmetry of galaxy optical brightness with respect to core/AGN \citep{Jacobs2014}.
The optical structure of the ICRF sources may be also complex; thus, high-resolution optical structure may be more difficult to investigate
than a radio structure.
Consequently, the accuracy of the ICRF--GCRF link might be limited to $\sim$0.5~mas \citep{Zacharias2014},
which is significantly worse than desirable given that the internal accuracy of ICRF and GCRF are at a level of a few tens of $\mu$as.

Using radio stars in a manner similar to that in the linking of {\it Hipparcos} Celestial Reference System (HCRF) to ICRF is another possibility
not yet considered for the alignment of GCRF to ICRF.
\citet{Froeschle1982} discussed the general principles for connecting optical and radio frames in view of the future {\it Hipparcos} mission
based on the VLBI observations of radio stars observed by {\it Hipparcos} in optics and connecting quasars to the {\it Hipparcos} stars
by the Space Telescope observations.
The authors concluded that using VLBI radio stars observations is the most powerful method among other considered in their study.
A numerical simulation showed that the connection between two frames can be obtained with an error of 1--4~mas
if the {\it Hipparcos} position error is 1--9~mas (depending on the optical brightness of the star) and VLBI position error is 2--5~mas.
Currently, with the GCRF--ICRF link, the accuracy of optical ({\it Gaia}) observations is expected to be about two order better
than that of {\it Hipparcos}.
The accuracy of the VLBI observations has also been substantially improved over the past two decades. 
Therefore, such a simulation should be revisited, which is the primary goal of this work.

For the final link of HCRF to ICRF, several methods have been used \citep{Kovalevsky1997}.
Among all the methods, the VLBI observation of radio stars has been considered the most accurate method and given the highest weight.
The results of the VLBI observations of 12 radio stars conducted between 1984 and 1994 allowed to determine the orientation angles between HCRF and ICRF
with an error of approximately 0.5~mas.

Several programmes for the astrometric observations of radio stars were conducted from the 1980s to the early 2000s
\citep{Lestrade1986,Garrington1995,Lestrade1995,Lestrade1999,Boboltz2003,Johnston2003,Boboltz2007}.
In the most recent work by \citet{Boboltz2007},
the VLBI positions of 46 radio stars with radio flux densities within the range of 1--10~mJy in the quiet state were determined.
The observations were made on the narrow network consisting of the Very Large Array (VLA) and the Pie Town Very Long Baseline Array VLBA antenna
located in $\sim$50~km from VLA.
The positions of radio stars were obtained using phase referencing of the stars to ICRF sources, with average error of $\sim$10~mas.
The orientation angles between the HCRF and ICRF frames were estimated with an error of $\sim$2.7~mas.

A review of the previous astrometric VLBI observations of radio stars shows that these observations were seldom and were mostly conducted
on narrow networks.
This is the most probable reasons why the accuracy of the radio star positions was relatively poor, and the errors in the obtained orientation
angles between the optical and radio frames from the observations were relatively high.
By contrast, properly scheduled radio star observations on regional or global VLBI networks could provide significantly lower position errors;
consequently, the accuracy of the link between the two frames could be enhanced.
In this study, a simulation is performed to determine the orientation angles between ICRF and GCRF by using radio stars.


\section{Modelling}
\label{sect:modeling}

The mutual rotation of the two frames is described by the following basic model using the three orientation angles $A_1$, $A_2$ and $A_3$:
\begin{equation}
\begin{array}{rcl}
\Delta\alpha & = & \phantom{-}A_1\cos\alpha\tan\delta + A_2\sin\alpha\tan\delta - A_3 \,, \\[1ex]
\Delta\delta & = & -A_1\sin\alpha + A_2\cos\alpha \,,
\end{array}
\label{eq:rotation_tanDE}
\end{equation}
where $\Delta\alpha$ and $\Delta\delta$ are the differences in the object coordinates in the two catalogues in right ascension and declination,
respectively.
These orientation angles are computed using the positions of common objects in the two compared frames.
In the present study, the common objects are the radio stars observed at both radio (VLBI) and optical ({\it Gaia}) wavelengths.

\begin{figure}
\centering
\resizebox{\hsize}{!}{\includegraphics[clip]{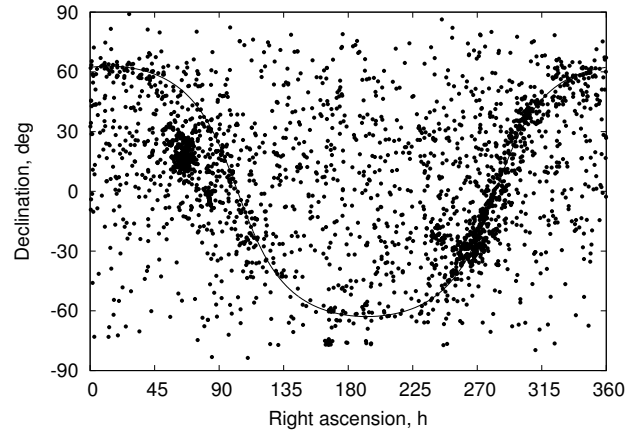}}
\caption{Distribution of the W95 radio stars over the sky. The Galactic equator is shown by the solid line.}
\label{fig:W95_sky}
\end{figure}

The radio stars catalogue by \citet{Wendker1995}, referred hereafter to as W95, is used for modelling.
It contains 3021 objects whose radio and optical brightness information are provided.
Approximately 3000 radio stars in W95 are brighter than 20~mag at optical wavelengths; thus, they are suitable for {\it Gaia} observations.
The distribution of the W95 stars over the sky is shown in Fig.~\ref{fig:W95_sky}.

\begin{figure}
\centering
\resizebox{\hsize}{!}{\includegraphics[clip]{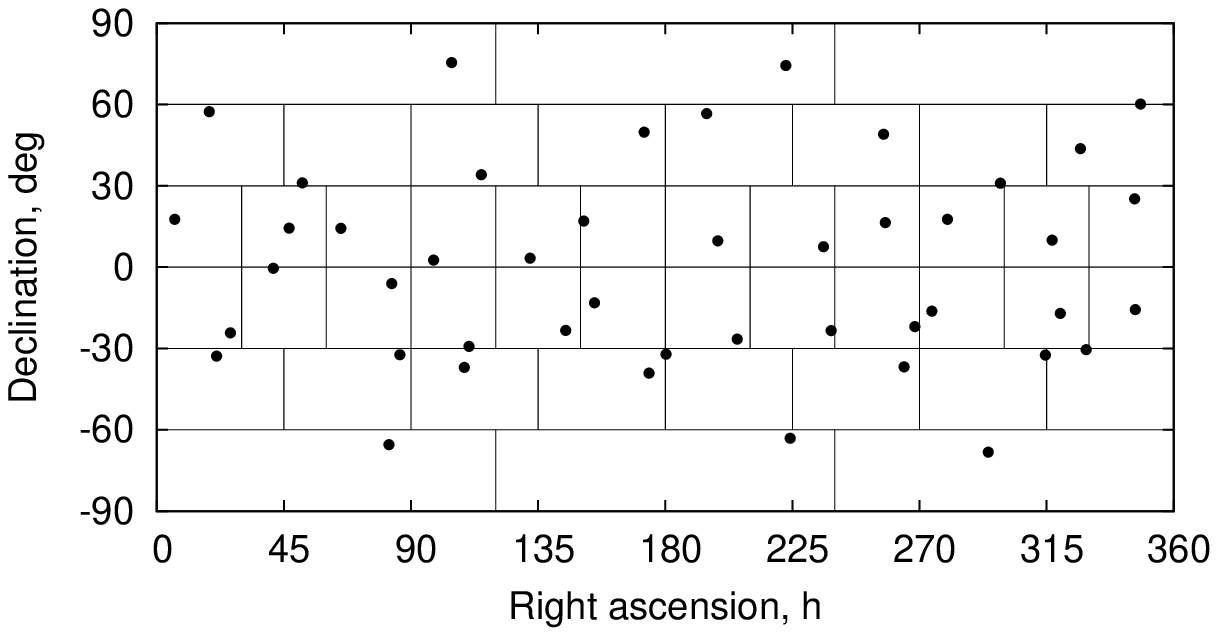}}
\resizebox{\hsize}{!}{\includegraphics[clip]{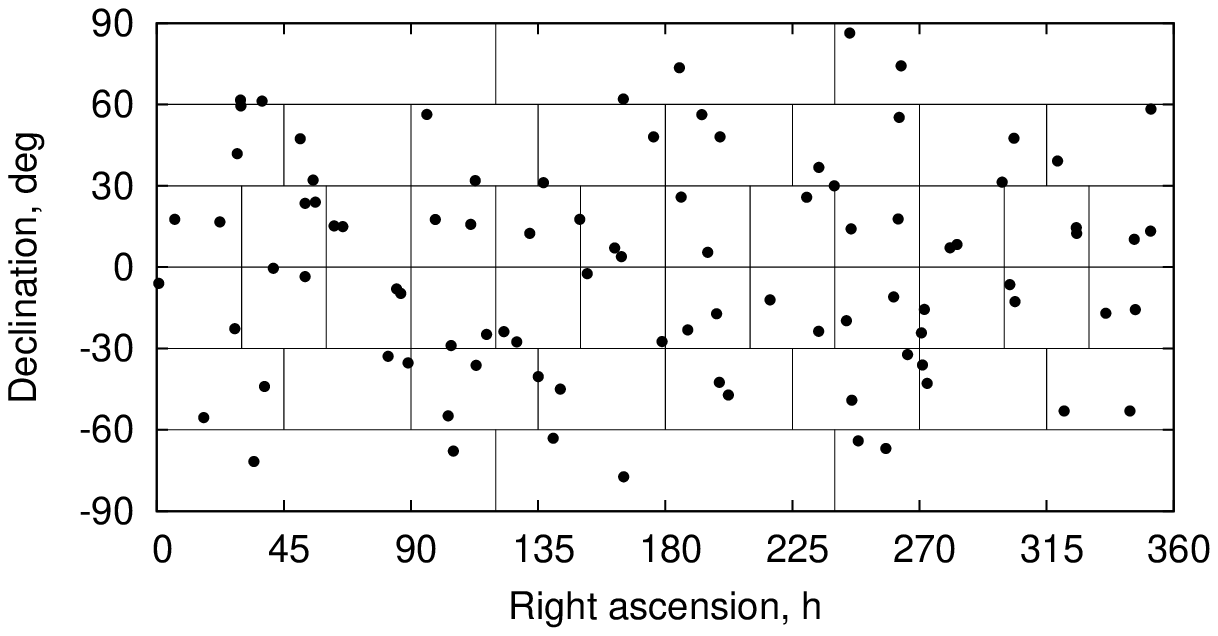}}
\resizebox{\hsize}{!}{\includegraphics[clip]{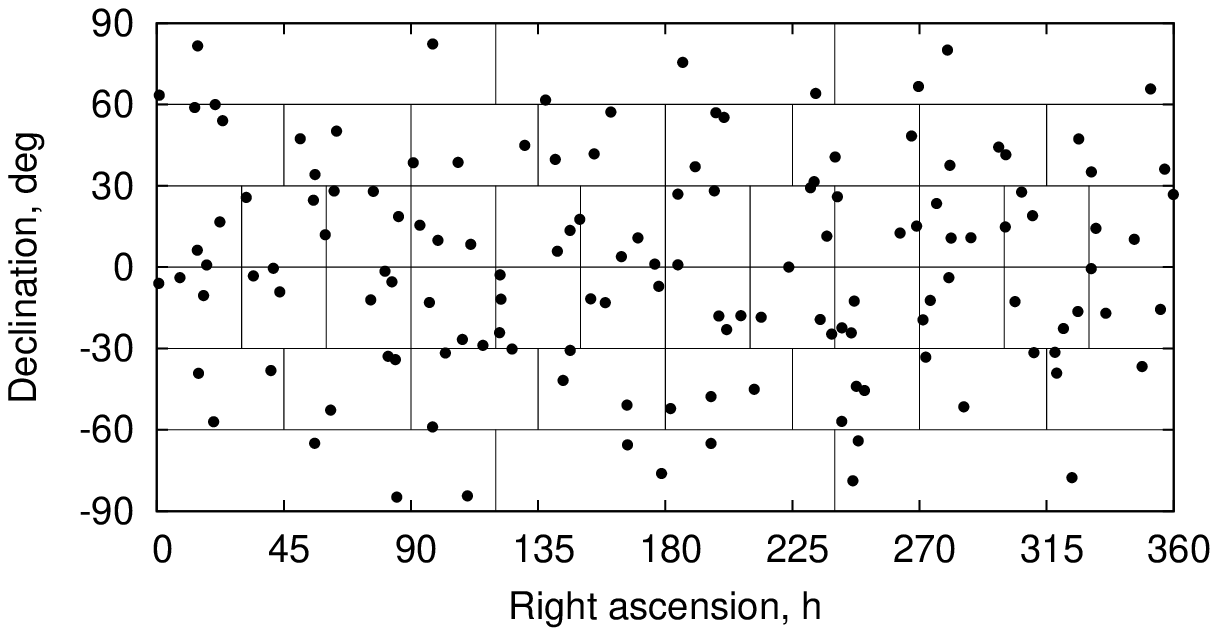}}
\caption{Example of the distribution of radio stars over 46 cells for random samples of 46, 92 and 138 radio stars (top to bottom).}
\label{fig:sky_stars_selections}
\end{figure}

For the estimation of the dependence of the orientation angle errors on the number of radio stars used, a Monte Carlo simulation with random
samples of radio stars was applied.
For a given number of radio stars, the errors in the orientation angles, as well as the correlation between them, are at a minimum
when the objects are randomly distributed over the sky.
Evidently, a simple random selection from W95 cannot provide satisfactory results because of the considerably uneven distribution
of the W95 objects, as shown in Fig.~\ref{fig:W95_sky}.
To provide a nearly uniform distribution of the selected radio stars over the sky, the celestial sphere is divided
into 46 cells of approximately 900 deg$^2$ each, as shown in Fig~\ref{fig:sky_stars_selections}.
Subsequently, one, two, or three radio stars were randomly selected from W95 in each cell.

For further modelling, the expected errors of the VLBI- and {\it Gaia}-derived positions should be estimated.
On the basis of astrometric VLBI experience, the average VLBI position errors may be in the range of 0.1--4~mas.
However, the values of these errors heavily depend on the radio flux density and number of observations, as discussed in Section~\ref{sect:discussion}. 
According to \citet{Bruijne2014}, the average errors in the {\it Gaia} positions are expected to be within the range of 5--400~$\mu$as,
depending on the optical brightness.

For each set of radio stars, their W95 coordinates were distorted by introducing random normally distributed errors to generate two mock catalogues. 
The first catalogue was computed using the VLBI position error $\sigma_{\mathrm V}$,
which was used as the standard deviation of the normal distribution to generate the source coordinate error added to the W95 position.
The second catalogue was computed using a similar method using the {\it Gaia} position error $\sigma_{\mathrm G}$.
Subsequently, the orientation angles between the two catalogues and their uncertainties, $\sigma_A$, were computed. 
In the computations, different $\sigma_{\mathrm V}$ and $\sigma_{\mathrm G}$ values, which are listed in Table~\ref{tab:angle_error}, were used.
It should be noted that $\sigma$ also includes a possible proper motion errors, which increase with the difference between the mean epochs of
VLBI and {\it Gaia} observations.
Given that radio stars are mainly optically bright objects, the proper motion errors are expected to be a few $\mu$as yr$^{-1}$ \citep{Bruijne2014}.

For the Monte Carlo simulation, 100,000 random samples of radio stars were generated.
The average uncertainties of the three orientation angles are shown in Table~\ref{tab:angle_error}.
The results show that the error in the orientation angles is mainly dominated by the VLBI position error except in the most optimistic scenarios.

\begin{table}
\begin{center}
\caption{Simulated 1-$\sigma$ uncertainty in the orientation angles between GCRF and ICRF. Unit: $\mu$as.}
\label{tab:angle_error}
\begin{tabular}{rrrrr}
\hline
$\sigma_{\mathrm V}$ & $\sigma_{\mathrm G}$ & \multicolumn{3}{c}{$\sigma_A$} \\
      &     & \multicolumn{3}{c}{Radio stars} \\
      &     &  46 &  92 & 138 \\
\hline
  100 &   5 &  18 &  13 &  10 \\
      &  10 &  18 &  13 &  10 \\
      &  25 &  18 &  13 &  11 \\
      &  50 &  20 &  14 &  12 \\
      & 100 &  25 &  18 &  15 \\
      & 200 &  40 &  28 &  23 \\
      & 400 &  73 &  52 &  42 \\[1ex]

  250 &   5 &  45 &  32 &  26 \\
      &  10 &  45 &  32 &  26 \\
      &  25 &  45 &  32 &  26 \\
      &  50 &  45 &  32 &  26 \\
      & 100 &  48 &  34 &  28 \\
      & 200 &  57 &  40 &  33 \\
      & 400 &  84 &  59 &  49 \\[1ex]

  500 &   5 &  89 &  63 &  51 \\
      &  10 &  89 &  63 &  51 \\
      &  25 &  89 &  63 &  52 \\
      &  50 &  89 &  63 &  52 \\
      & 100 &  91 &  64 &  52 \\
      & 200 &  96 &  68 &  55 \\
      & 400 & 114 &  81 &  66 \\[1ex]

 1000 &   5 & 178 & 126 & 103 \\
      &  10 & 178 & 126 & 103 \\
      &  25 & 178 & 126 & 103 \\
      &  50 & 178 & 126 & 103 \\
      & 100 & 179 & 127 & 103 \\
      & 200 & 181 & 128 & 105 \\
      & 400 & 192 & 136 & 111 \\[1ex]

 2000 &   5 & 356 & 252 & 206 \\
      &  10 & 356 & 252 & 206 \\
      &  25 & 356 & 252 & 206 \\
      &  50 & 356 & 252 & 206 \\
      & 100 & 356 & 252 & 206 \\
      & 200 & 358 & 253 & 207 \\
      & 400 & 362 & 257 & 210 \\[1ex]

 4000 &   5 & 712 & 504 & 411 \\
      &  10 & 712 & 504 & 411 \\
      &  25 & 712 & 504 & 411 \\
      &  50 & 712 & 504 & 411 \\
      & 100 & 712 & 504 & 411 \\
      & 200 & 712 & 504 & 412 \\
      & 400 & 715 & 506 & 413 \\[1ex]
\hline
\end{tabular}
\end{center}
{\bf Notes:} $\sigma_{\mathrm V}$ is the VLBI position uncertainty, $\sigma_{\mathrm G}$ is the {\it Gaia} position uncertainty,
  $\sigma_A$ is the resulting orientation angles uncertainty.
\end{table}

\begin{figure}
\centering
\resizebox{\hsize}{!}{\includegraphics[clip]{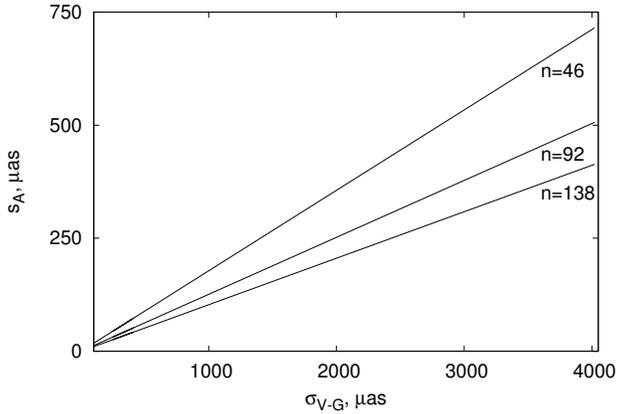}}
\caption{Dependence of the error in the orientation angles on the error in the VLBI--{\it Gaia} position difference $\sigma_{\mathrm V-G}$ for different
number of stars ($n$).}
\label{fig:sobs_sa}
\end{figure}

Fig.~\ref{fig:sobs_sa} shows the dependence of the error in the orientation angles $\sigma_A$ on the error in the VLBI--{\it Gaia}
position difference $\sigma_{\mathrm V-G} = \sqrt{\sigma_{\mathrm V}^2+\sigma_{\mathrm G}^2}$.
This dependence can be well described by the following linear model:
\begin{equation}
\sigma_A = \left\{\begin{array}{ll}
0.178 \; \sigma_{\mathrm V-G}, & n=46 \,, \\
0.126 \; \sigma_{\mathrm V-G}, & n=92 \,, \\
0.103 \; \sigma_{\mathrm V-G}, & n=138 \,.
\end{array}
\right.
\label{eq:sobs_sa}
\end{equation}

To find an analytical expression for the factor in equation (\ref{eq:sobs_sa}), a supplement simulation was conducted with the number
of stars being, 184, 230 and 276, with 4, 5 and 6 stars in each cell.
For the number of stars in the range from 46 to 276, the dependence of the error in the orientation angles on the $\sigma_{\mathrm V-G}$ and $n$ can be
approximated by
\begin{equation}
\sigma_A = (n^{-0.432}-0.015) \; \sigma_{\mathrm V-G} \,. \\
\label{eq:sobs_sa_final}
\end{equation}

It is interesting to compare the modelled $\sigma_A$ values with the actual error obtained for the HCRF--ICRF link.
For this link, the observations of 12 radio stars were used with a median VLBI position uncertainty, $\sigma_{\mathrm V}$, of 1.5~mas \citep{Kovalevsky1997}.
If the {\it Hipparcos} position uncertainty for the radio stars is 1~mas, the $\sigma_A$ value calculated by equation (\ref{eq:sobs_sa_final})
is 0.6~mas, which is in good agreement with the value of 0.5~mas obtained in practice by \citet{Kovalevsky1997}.


\section{Discussion}
\label{sect:discussion}

The simulation results presented in Table\ref{tab:angle_error} provide only a general impression of the possible accuracy
of the ICRF-GCRF link realized using radio stars.
Several practical issues should be further considered.

\begin{figure}
\centering
\resizebox{\hsize}{!}{\includegraphics[clip]{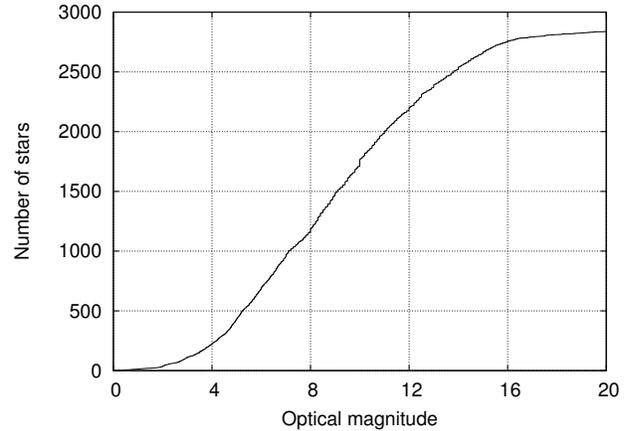}}
\caption{Distribution of the W95 radio stars over the optical magnitudes in the $B$ to $R$ range.
  The plot shows the number of stars with brightness equal to the given magnitude or brighter than the given magnitude.}
\label{fig:mag_n_W95}
\end{figure}

The expected errors in the {\it Gaia} positions depends on the optical magnitude of the star.
They range from 5~$\mu$as for the objects with magnitudes of $\le$12~mag to $\sim$450~$\mu$as for the objects with a magnitude
of 20~mag \citep{Bruijne2014}.
Fig.~\ref{fig:mag_n_W95} shows the distribution of the W95 radio stars over the optical magnitudes in the $B$ to $I$ range.
W95 contains one visual or IR magnitude for each star over a wide range from $\sim$0.45$\mu$ ($B$ band) to 11$\mu$ (LWIR band).
In Fig.~\ref{fig:mag_n_W95}, 2841 stars with given magnitudes in the $B$ to $R$ range are used.
No attempt has been made to reduce all the magnitudes to a single band because these tentative magnitudes are sufficient for the current simulation.
However, it can be mentioned that the average $B-R$ colour index for the astrometric radio sources is 0.9 \citep{malkin2016d}.
Most of the radio stars in W95 have magnitude $<$16 mag, which corresponds to the {\it Gaia} position error of $<$~30~$\mu$as.
Approximately 75 per cent of radio stars are brighter than 12~mag, and the expected accuracy of their positions is 5~$\mu$as \citep{Bruijne2014}.
Therefore, the `optimistic' lines in Table~\ref{tab:angle_error} related to the {\it Gaia} position accuracy can be considered.
However, doing so will generate a practical difference only in the case of VLBI positions with accuracy levels of $\le$0.5~mas.

\begin{figure}
\centering
\resizebox{\hsize}{!}{\includegraphics[clip]{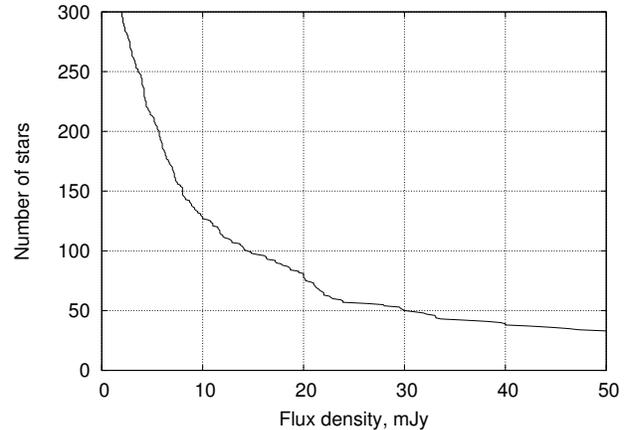}}
\caption{Distribution of weak W95 radio stars over the flux densities.
   The plot shows the number of radio stars having flux densities equal to or larger than the given value.}
\label{fig:flux_n_W95}
\end{figure}

The accuracy of VLBI observations depends on the flux density of the radio source.
Weak sources cannot be detected at all.
Thus, the flux densities of the radio stars should be determined to assess their suitability for astrometric VLBI observations.
Fig.~\ref{fig:flux_n_W95} shows the distribution of the W95 radio stars with the flux density of 50 mJy and less.
Only the W95 objects with explicitly given flux densities were used for this plot.
A flux density was not suggested if only the upper limit is given in W95.

The next question is whether faint radio stars with a flux density of several mJy can be detected on the VLBI baselines.
Experience has shown that this is possible.
As mentioned above, radio stars with flux densities of only a few mJy were successfully observed in a previous observing programme \citep{Boboltz2007}.

\citet{Bourda2010} investigated the minimum detectable flux density on 10 baselines with the use of 5 large- and medium-sized European antennae,
namely, Effelsberg (100~m), Robledo (70~m), Medicina (32~m), Noto (32~m) and Onsala (20~m). 
Observations were performed at the S and X bands with a scan length of 5~min and a registration rate of 1~Gbps.
SNRs of $\ge$ 7 were achieved for radio sources with a correlated flux densities of 0.7/3~mJy (at the X/S bands) on the baseline Effelsberg--Robledo,
and approximately 5/6~mJy on the baselines Robledo--Medicina and Robledo--Noto.
The minimum detectable flux densities at the X and S bands were nearly the same for all the baselines not involving the Effelsberg antenna,
which had limited sensitivity at the S band during the experiment \citep{Bourda2010}.

\begin{figure}
\centering
\resizebox{\hsize}{!}{\includegraphics[clip]{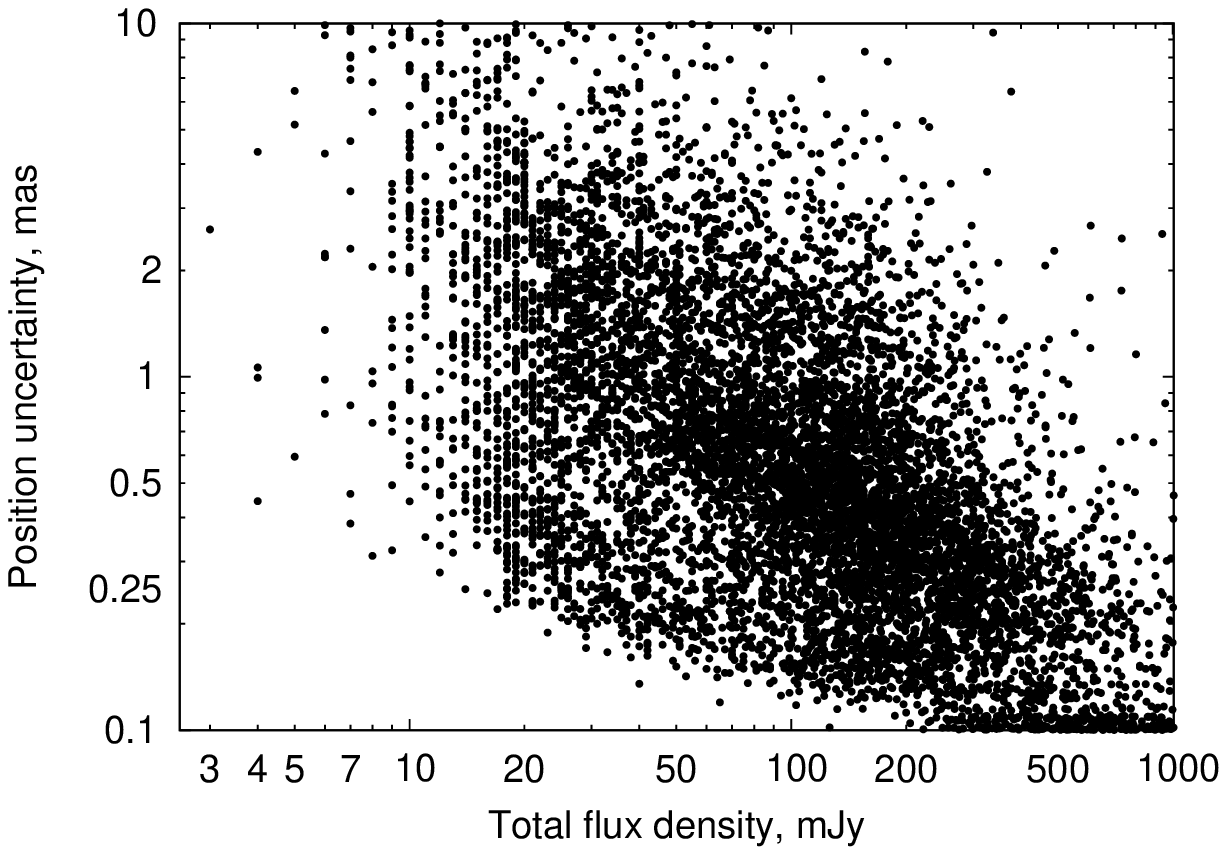}}
\resizebox{\hsize}{!}{\includegraphics[clip]{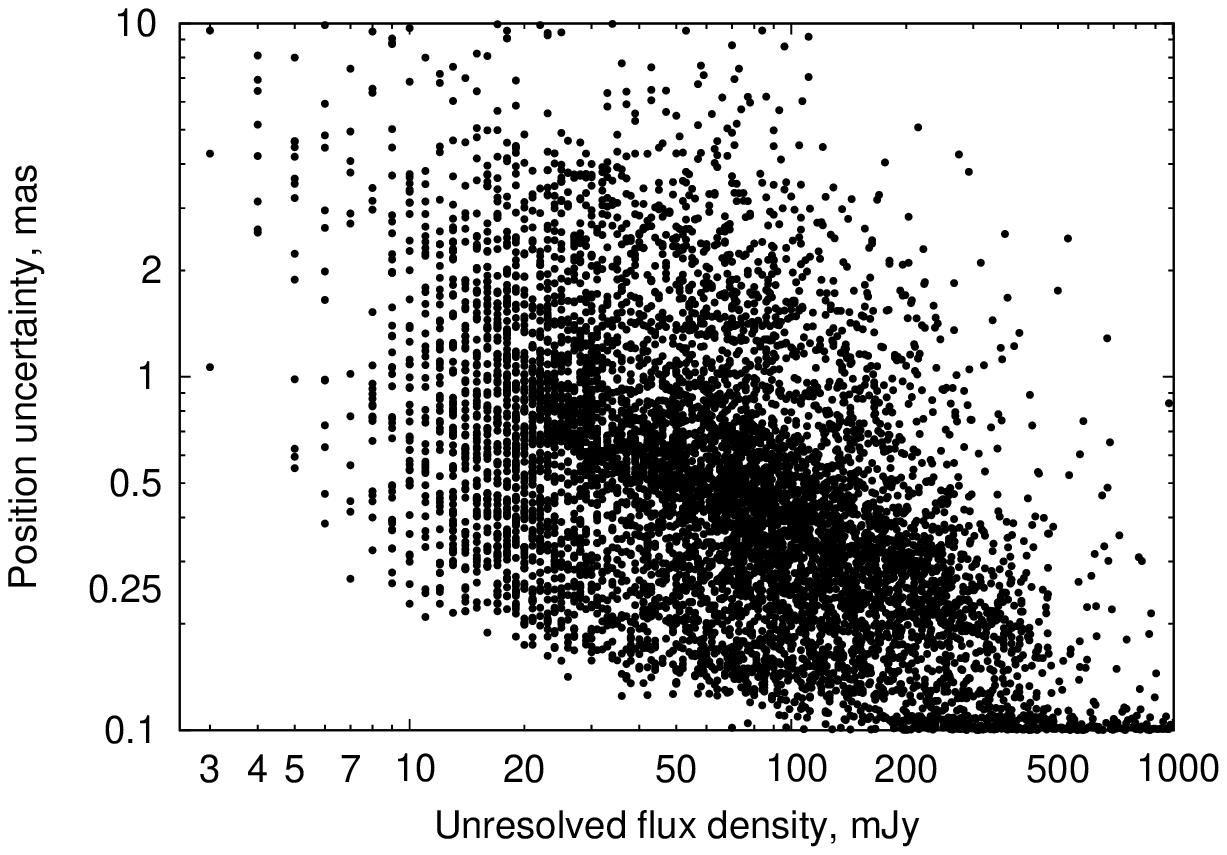}}
\caption{Radio source position uncertainty vs. total (top) and unresolved (bottom) X-band flux densities according to the RFC2016a catalogue.}
\label{fig:sigma_flux}
\end{figure}

The dependence of the VLBI position uncertainty on the flux density according to the RFC2016a
catalogue\footnote{http://astrogeo.org/vlbi/solutions/rfc\_2016a/},
which is currently the most comprehensive catalogue of astrometric radio sources whose positions and flux densities are provided,
is depicted in Fig.~\ref{fig:sigma_flux} for both total and unresolved flux densities.
The RFC catalogue provides two estimates of the correlated source flux densities, namely, total flux density and unresolved flux density.
The total (integrated) flux density is computed as the sum of the CLEAN or other high-frequency spatial components,
and the unresolved flux density is estimated for the sources with unresolved structure.
Details on the computation of the source flux density in the RFC catalogue can be found in \citet{Petrov2008}.
Although the plots of both total and unresolved flux densities appear similar, the points positioned in close proximity on both plots
are generally unrelated to the same source because the total and unresolved source flux densities may differ by a factor of ten and even more
depending on source compactness, telescopes sensitivity, network size and configuration.
These estimates can hardly be modelled reliably; thus, they must be obtained from observations.
Therefore, the two plots in Fig.~\ref{fig:sigma_flux} complement each other and consequently provide more comprehensive flux measurement information.
The position uncertainties shown in the plots are computed as $(\sigma_\alpha \cos\delta + \sigma_\delta)/2$,
where $\sigma_\alpha$ and $\sigma_\delta$ are the source position uncertainties in right ascension and declination, respectively.
One can see that the RFC2016a catalogue contains positions of several faint sources with flux densities of only a few mJy.

\begin{figure}
\centering
\resizebox{\hsize}{!}{\includegraphics[clip]{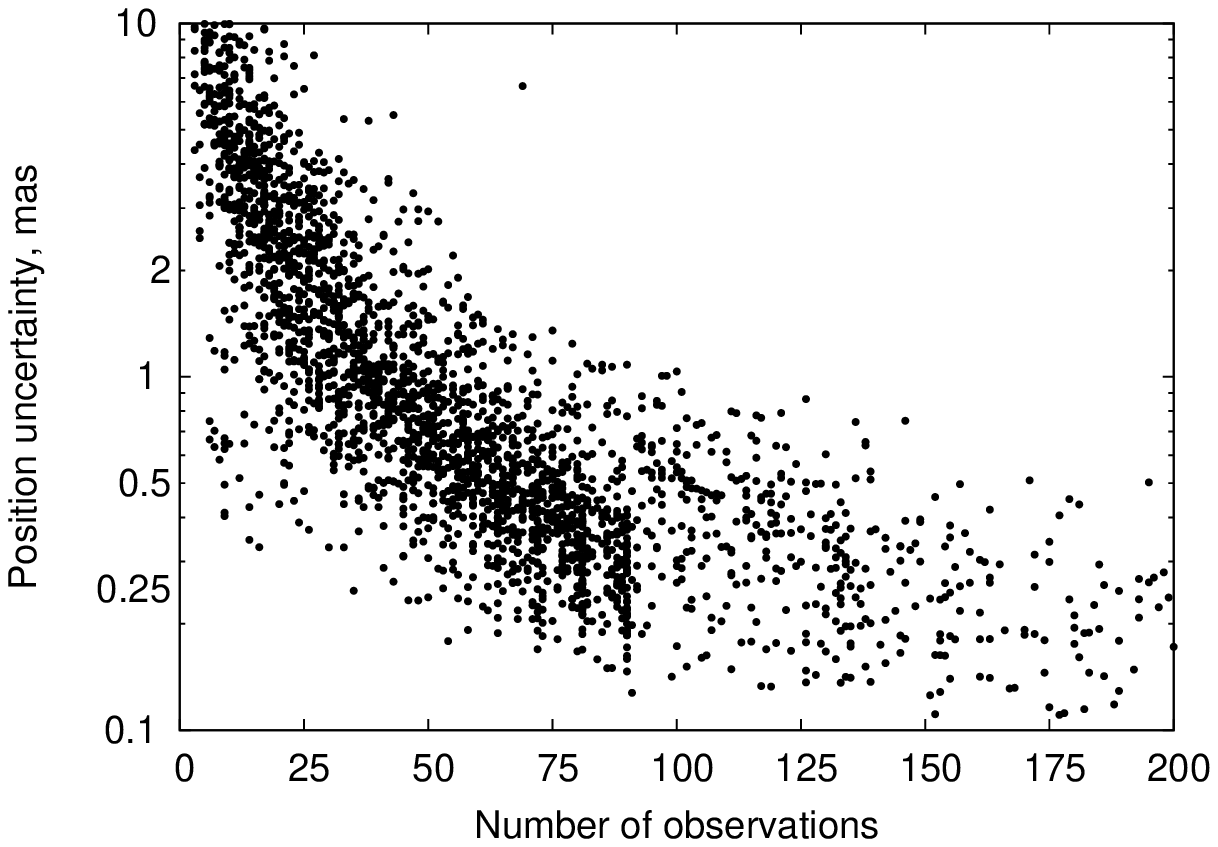}}
\resizebox{\hsize}{!}{\includegraphics[clip]{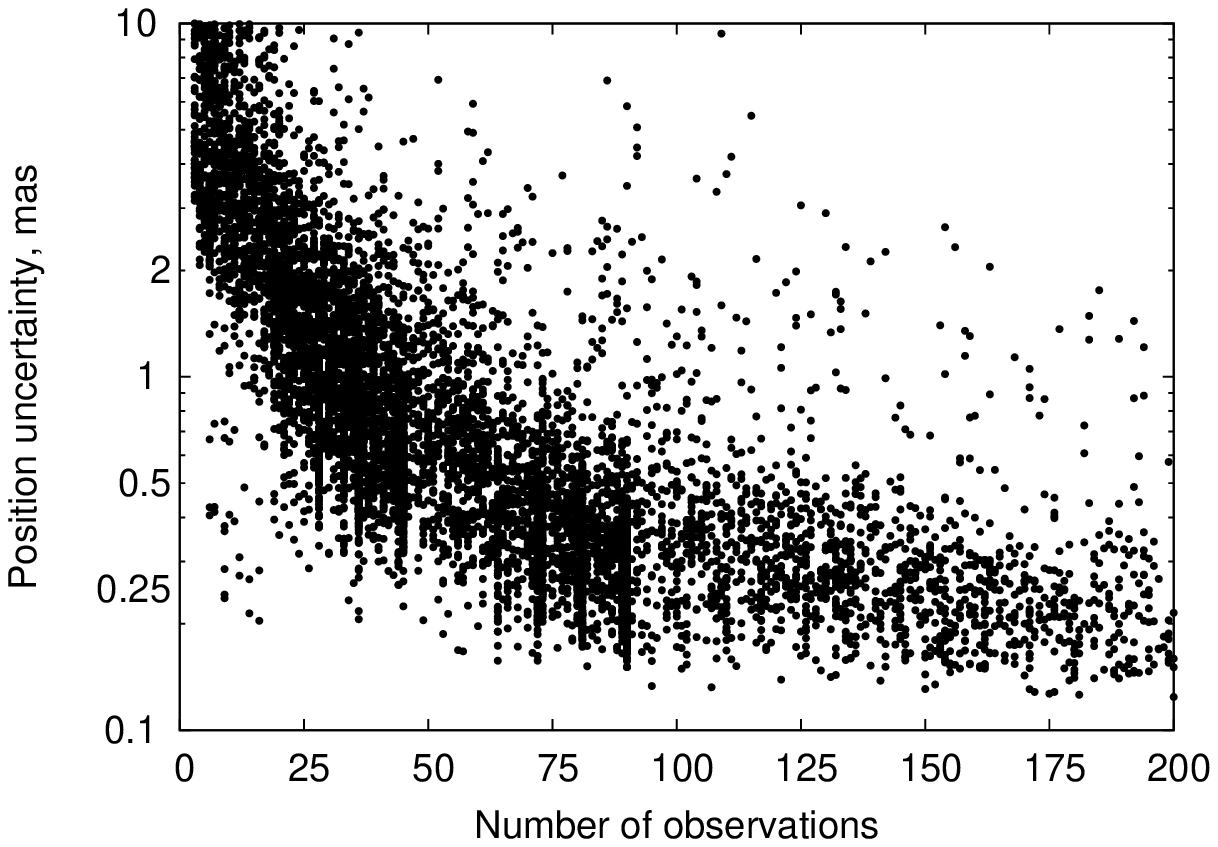}}
\caption{Radio source position uncertainty vs. the number of observations for ICRF2 (top) and RFC2016a (bottom) catalogues.}
\label{fig:sigma_nobs}
\end{figure}

Therefore, radio stars with flux densities of 4--5~mJy or higher can be observed by astrometric VLBI,
provided large antennae are included in the network.
W95 contains 214 radio stars with flux densities of $\ge$~5~mJy at the S/X bands, and 129 of them have the flux densities of $\ge$~10~mJy.
The actual accuracy levels of the radio star positions obtained from these observations depend on several factors, such as schedule,
number of observations, antennae used, network geometry, and registration mode. 
With regard to the last factor, modern VLBI backend systems can accommodate registration rates of 2~Gbps and higher, allowing for more reliable
observations of faint sources.

Realistic errors in the VLBI positions of radio stars should also be estimated.
An investigation of the dependence of the RFC2016a position error on the flux density of a source (Fig.~\ref{fig:sigma_flux})
shows that even for weak sources each with a flux density of 5~mJy, a position error below 1 mas~can be obtained. 
Many sources with flux densities of 7--10~mJy have the position errors in the range of 0.25--0.5~mas.

VLBI position accuracy also depends on the number of observations, as shown in Fig.~\ref{fig:sigma_nobs}.
Approximately 50 observations are required to obtain a VLBI position error of 1~mas, and approximately 100 observations are needed
to obtain a VLBI position error below~0.5 mas.

As shown in Table~\ref{tab:angle_error}, observing more than 100 radio stars with an average VLBI position accuracy of 1~mas allows
for the derivation of the orientation angles between GCRF and ICRF with an error of approximately 0.1~mas. 
A twofold lower error in the orientation angles can be obtained with a VLBI position accuracy of 0.5~mas.

Experience with {\it Hipparcos} has shown that the HCRF--ICRF link can be obtained using only 12 radio stars with an error of approximately 0.5~mas.
Significantly improved optical ({\it Gaia} vs. {\it Hipparcos}) and VLBI (with improvements in VLBI technology) observation accuracy levels,
as well as the use of a considerably higher number of radio stars allow for the significant reduction in the error in the link between radio
and optical frames.

However, some accuracy-limiting factors in radio star observations should be considered.

First, many radio stars comprise double or multiple systems.
Thus, their orbital motions should be taken into account \citep{Kovalevsky1997,Lestrade1999}.
The accurate {\it Gaia}-derived orbits can be used for this purpose.

Second, radio stars may have complex and variable structures, which might cause a measurable jitter in the star position and a bias between
the optical and radio positions.
\citet{Lestrade1995} estimated the impact of radio star structure and possible variations in the radio star emitting centre to be within
the error budget of approximately 0.5 mas.
\citet{Lestrade1999} found that the structure-induced systematic errors in the VLBI positions of 12 stars ranged from 0.07~mas to~0.54 mas,
with a median value of 0.18~mas, which is close to the lower limit of the VLBI position error used in the simulation discussed
in Section~\ref{sect:modeling}.
Provided several tens of radio stars have been observed, this factor should not significantly impact the errors in the orientation
angles except for the most optimistic cases presented in the first lines of Table~\ref{tab:angle_error}. 
Nevertheless, the investigation and proper modelling of the structures of radio star emitting regions is a worthy endeavour for
the future improvement of the astrometry of radio stars.

Finally, the best strategy for using radio stars to link the GCRF to ICRF may be worth discussing. 
The radio stars positions used for this link must be determined in the ICRF.
Since the radio stars are not ICRF objects, their VLBI observations should be organized in a special way.
In the previous works, radio stars positions were tied to the ICRF through phase referencing to close ICRF sources.
An alternative strategy can be considered when the radio stars are included in a VLBI observing schedule together with the ICRF sources.
Then the radio stars positions can be obtained in a global VLBI solution.
Such a strategy can be profitable as it allows more flexible stars selection (no nearby ICRF source is needed) and may reduce systematic errors
of the result.


\section{Conclusion}
\label{sect:conclusion}

The VLBI observations of radio stars, which are also suitable for observations with Gaia, is a prospective method to link GCRF to ICRF.
This method has been successfully applied for the past two decades to align HCRF to ICRF with accuracy of a few tenths of mas.
With the improved accuracy of modern radio VLBI and that of optical {\it Gaia} observations, the accuracy of the GCRF--ICRF link
can be significantly enhanced.

The simulation conducted in this study has shown that radio star observations can theoretically allow for the realization of the link
between GCRF and ICRF with the errors in orientation angles  at a level of 10$\mu$as. 
Although achieving such an accuracy is tempting, it is overly optimistic in view of the possible unmodelled errors of VLBI and optical
observations and the unrealistically considerable effort required of the VLBI community to allocate sufficient observation time,
particularly if large antennae are involved.
However, a realistic properly organized VLBI observing programme for radio stars can allow for the realization of the GCRF-ICRF link
with an accuracy of ~0.1~mas. 
Thus, this method can provide a valuable contribution to the improvement of the GCRF--ICRF link.

\section*{Acknowledgements}

This work was partly supported by the Russian Government program of Competitive Growth of Kazan Federal University.

This research has made use of NASA's Astrophysics Data System.

The author is grateful to the anonymous referee for constructive comments and useful suggestions.

\bibliography{radiostars_v3}
\bibliographystyle{mn}

\end{document}